# Strain relaxation in GaN grown on vicinal 4H-SiC(0001) substrates

J. Pernot and E. Bustarret
Institut NEEL, CNRS & Université Joseph Fourier BP166 F-38042 Grenoble Cedex 9, France

M. Rudziński, P. R. Hageman, P. K. Larsen
Applied Materials Science, Institute for Molecules and Materials, Radboud University, Toernooiveld 1, 6525 ED Nijmegen, The Netherlands

**Abstract**
The strain of GaN layers grown by Metal Organic Chemical Vapor Deposition (MOCVD) on three vicinal 4H-SiC substrates (0, 3.4 and 8 offcut from [0001] towards [11-20] axis) is investigated by X-ray Diffraction (XRD), Raman Scattering and Cathodoluminescence (CL). The strain relaxation mechanisms are analyzed for each miscut angle. At a microscopic scale, the GaN layer grown on on-axis substrate has a slight and homogeneous tensile in-plane stress due to a uniform distribution of threading dislocations over the whole surface. The GaN layers grown on miscut substrates presented cracks, separating areas which have a stronger tensile in-plane stress but a more elastic strain. The plastic relaxation mechanisms involved in these layers are attributed to the step flow growth on misoriented surfaces (dislocations and stacking faults) and to the macroscopical plastic release of additional thermoelastic stress upon cooling down (cracks).

## I.   Introduction

GaN is an attractive semiconductor because of its intrinsic properties for electronic (high power, high frequency and high temperature) and optoelectronic (UV) applications. However, up to now, no substrate combining large size and high quality is commercially available. GaN-based devices are therefore developed using hetero-structures on a host substrate having interesting structural, thermal and electrical properties. Among such templates, silicon carbide is a good candidate with a high thermal conductivity and a relatively large substrate sizes available (conducting and semi-insulating substrates). Its lattice parameters ($a_{4H-SiC}$= 3.08051 Å and $c_{4H-SiC}$= 2.52120 Å for 4H[1]) are close to those of GaN ($a_{GaN}$= 3.1890 Å and $c_{GaN}$= 2.5932 Å for hexagonal GaN[2]). Combined with a larger thermal expansion coefficient for GaN ($5.59 \times 10^{-6}$) than for SiC ($3.7 \times 10^{-6}$ for 4H), this yields GaN epilayers generally under tensile in-plane strain[3] and leads to the formation of a high density of linear and planar defects such as dislocations, stacking faults or cracks. At present, one of the most important challenges is to determine the best growth conditions to obtain a thin GaN epilayer fully unstrained with a high crystal quality. A promising technique is the vicinal surface epitaxy using a misoriented substrate[4]. This technique has the advantage to grow the GaN epilayer with a step flow mechanism limiting the generation of threading dislocations[5]. Recently, Huang et al. proposed to control the strain relaxation with a new mechanism, using an optimized value of the miscut angle of the SiC substrate[6].
In this work, we will study the influence of the miscut angle of 4H-SiC substrate on strain relaxation in the GaN epilayer. First, the experimental details will be described.

Then, the results of X-Ray diffraction as well as microRaman and cathodoluminescence spectroscopies are used in order to evaluate the strain state of the layers grown on differently oriented 4H-SiC. Finally, the strain relaxation mechanisms are discussed before a brief conclusion.

## II. Experimental details

We focus on three GaN/AlGaN/4H-SiC samples grown under identical growth conditions (side by side in the low pressure MOCVD reactor). Trimethylgallium (TMGa), trimethylaluminium (TMAl), and ammonia ($NH_3$) were used as precursors and $H_2$ as a carrier gas. A nucleation layer consisting of AlGaN with a low Al content ($\cong$ 1 at.%) and with a thickness of 120 nm was employed. The non-intentionally doped GaN layer was grown at identical conditions as the nucleation layer, 1170°C at 50 mbar total reactor pressure, and the growth time was set to yield about 1.8 μm thick GaN layers for all samples discussed in this paper. For the present experiments, we used nominally (0001)-oriented (on-axis) 4H-SiC substrates and substrates that were 3.4°, and 8° misoriented from (0001) towards [11$\bar{2}$0].

To check the structural quality of epilayers grown on these substrates, the samples have been examined by X-ray using the rocking curve mode using a Bruker D8 Discovery X-ray diffractometer with a Cu target ($\lambda$=1.54060 Å) and a 4-bounce Ge (022) monochromator.

Micro-Raman measurements were performed using a Jobin-Yvon Labram Infinity spectrometer fitted with an Olympus BX40 microscope and a liquid nitrogen-cooled CCD detector. A 2400 gr/mm grating led to a pixel spectral separation of 0.42 $cm^{-1}$. The 633 nm line of a HeNe laser served as excitation source. To achieve a spatial resolution of 2 $\mu m^3$, a ×100 objective was used with a confocal pinhole of 100 μm. The backscattering configuration was employed with both the incident and scattered light propagating along the $z$ direction (parallel to the [0001] direction of h-GaN) and analyzed for polarization along an arbitrary direction $x$, perpendicular to the [0001] direction of h-GaN, and undetermined in this plane. For samples with 3.4° and 8° off-axis toward [11$\bar{2}$0], the incident and scattered light are equally misoriented with respect to the [0001] direction.

Optical properties of the samples were assessed by cathodoluminescence (CL). The CL measurements were performed in a Scanning Electron Microscope (SEM) equipped with a He cooling stage at beam currents and energies varying respectively from 10 pA to 80 nA and from 20 to 30 keV. The temperature of the sample was between 12 K and 20 K. The emission spectra were analyzed using a 0.4m monochromator equipped with a UV-enhanced liquid nitrogen-cooled CCD camera. The spectral resolution was estimated to be 0.5 Å, which, in the energy range of interest, is equivalent to 0.5 meV. Depending on the resolution requirements, a 600gr/mm grating blazed at 300 nm and a 1800 gr/mm grating blazed at 500 nm were employed.

## III. Results and discussion
**A. X-Ray**
The X-ray rocking curves of our samples have been presented in previous works[7]. The structural quality of the GaN epilayer is decreasing at larger off-cut angles, in particular the Full width at Half Maximum (FWHM) of the (105) peak increases dramatically. The lattice parameters *a* and *c* are calculated using the standard formula taking into account the refraction index[8]. The normal ($\varepsilon_{zz}$) and in plane ($\varepsilon_{xx}$) components of the strain tensor are deduced using the strain free lattice parameters of GaN[2]: $a_0$=3.1890 Å and $c_0$=5.1864 Å. The ratio $R^B$= $\varepsilon_{zz}/\varepsilon_{xx}$ is evaluated for comparison with the value predicted by the elastic theory in the case of a pure biaxial stress $R^B$= - 0.53±10, according to the experimental data given by Polian et al.[9]. Results are summarized in Table 1. The in-plane strain is tensile for the three samples and increases with the misorientation angle of the substrate. Along c-axis direction, the normal strain is compressive, yielding a negative ratio $R^B$ in agreement with the elastic strain rules. The $R^B$ value of the 8° off axis sample is very close to the one corresponding to an ideal elastic relaxation of the stress. At the opposite, the ratio $R^B$ deviates from this value as the miscut angle decreases. These internal strain measurements have been confirmed by microRaman spectroscopy on the same layers.

| Sample | off axis angle (°) | Thickness (μm) | a (Å) | c (Å) | $\varepsilon_{xx}$ (%) | $\varepsilon_{zz}$ (%) | $R^B$ | $E_2$ (cm$^{-1}$) | $\sigma^B$ (GPa) | Cracks |
|---|---|---|---|---|---|---|---|---|---|---|
| #1 | 0 | 1.80 | 3.1930 ± 0.0005 | 5.18085 ± 0.0003 | +0.1270 ± 0.016 | -0.1070 ± 0.004 | -0.843 | 567.1 | -0.3745 ± 0.12 | No |
| #2 | 3.4 | 1.65 | 3.1980 ± 0.001 | 5.1776 ± 0.0003 | +0.2822 ± 0.029 | -0.1697 ± 0.007 | -0.601 | 565.9 | -0.8642 ± 0.28 | along the [0112], [0211] and [0121] axis. |
| #3 | 8 | 1.65 | 3.2005 ± 0.001 | 5.17675 ± 0.0003 | +0.3606 ± 0.029 | -0.1861 ± 0.006 | -0.516 | 564.9 | -1.2757 ± 0.42 | parallel along [1120] axis |

Table 1: Samples characteristics and strain data deduced from X-Ray and Raman measurements.

**B. Micro-Raman spectroscopy**

Micro-Raman scattering is a useful non destructive probe of the symmetry of a crystal and of the strain in epilayers at the micrometer scale. Hexagonal GaN grows in the wurtzite structure and belongs to the $C_{6v}$ (6mm) point group. Near **k=0**, group theory predicts four Raman active modes: one $A_1$, one $E_1$, and two $E_2$. The strong $E_2$ mode at 568 cm$^{-1}$ is allowed for the two backscattering configurations (parallel $z(x,x)\bar{z}$ and crossed $z(x,y)\bar{z}$). Its nonpolar character and its elevated scattering cross section make it the best phonon mode to evaluate the stress in GaN[3]. The figures 1 and 2 show the $E_2$ Raman mode spectra for the three samples. The presence of the TO mode of c-GaN at 554.5 cm$^{-1}$ in the 8° off-axis sample, indicated in Fig.1 by an arrow, confirms the presence of cubic inclusions in the 8° miscut angle sample as already observed by HRTEM[10]. The shift of the line position and the broadening of the $E_2$ mode in Fig. 2 indicate a tensile stress which increases with the miscut angle of the substrates. The Raman spectra have been fitted using Lorentzian lineshapes. The energy positions are reported in Table 1 together with the X-Ray data and in Fig. 3 as a function of $\varepsilon_{xx}$. The experiments are within the expected frequency range represented by the grey area deduced from the usual relation $\Delta\omega(E_2) = 2 a \varepsilon_{xx} + b \varepsilon_{zz}$, where $a$ and $b$ are the phonon deformation potentials. The phonon deformation potential values were recently determined by Demangeot et al.[11] and used here with their error bars ($a$= -850±177 cm$^{-1}$ and b= -963±220 cm$^{-1}$). The dependence of the shift $\Delta\omega$ versus $\varepsilon_{xx}$ is not linear in our case since $R^B$ is not a constant. Then, the biaxial stress $\sigma^B$ for each layer has been evaluated using the Raman biaxial pressure coefficient $K^B$ and the $E_2$ phonon shift measured for each sample[11]. It is important to notice that we have systematically located the laser spot far away from the cracks, observed in the 3.4 and 8 off axis samples. The present GaN layer grown on the on axis substrate is less strained than the thin GaN epilayers grown on 6H-SiC (0001) substrates studied in Ref. 12.

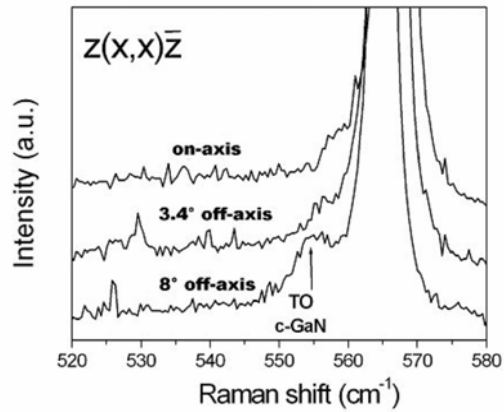

Fig.1

FIG. 1. Room temperature micro-Raman spectra of h-GaN epilayers deposited on 4H-SiC substrates with misorientation of 0°, 3.4° and 8° toward [11$\bar{2}$0]. Arrows show signature of c-GaN inclusions in 8° off-axis sample.

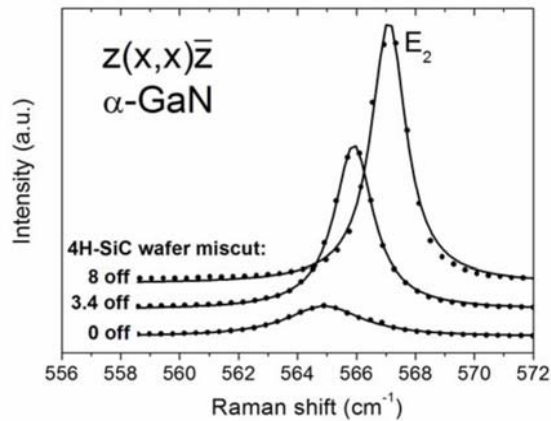

Fig.2

FIG. 2. Room temperature micro-Raman spectra of h-GaN epilayers deposited on 4H-SiC substrates with misorientation of 0°, 3.4° and 8° toward [11$\bar{2}$0]. Symbols are experimental data and full line a calculation using Lorentzian lineshapes.

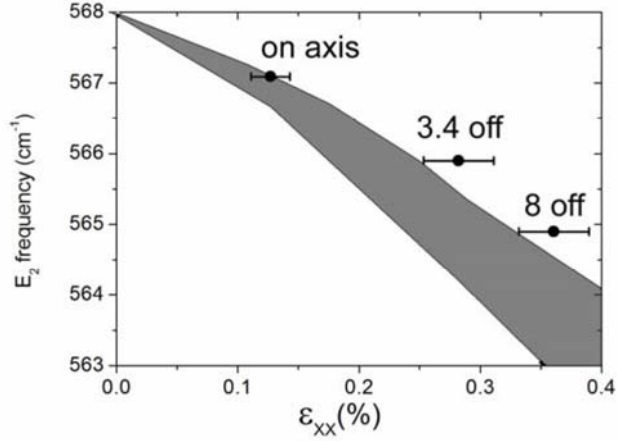

Fig.3

FIG. 3. $E_2$ mode frequency versus the in plane strain $\varepsilon_{xx}$ for the three samples with different substrate miscut angle. The gray area represents the expected dependence for a pure elastic deformation.

**C. Cathodoluminescence spectroscopy**

The low-temperature CL spectra of the hexagonal GaN layers grown on the differently oriented 4H-SiC substrates (0°, 3.4° and 8° off-axis) are shown in Fig. 4. The spectra have some common features in two spectral ranges. First, between 3.44 and 3.475 eV the donor bound exciton (DBE) emission lines are seen for all three substrate orientations, as well as their LO phonon replica red-shifted by 91-92 meV. Secondly, the donor-acceptor pair band (DAP) of h-GaN at 3.26 eV and its first (LO) and second (2LO) phonon replica, as assigned in Figs. 4. Extra features, indicated by arrows and star in Fig. 4, appear in the CL spectra of the 8° misoriented sample: *i)* a broad peak at ~3.39 eV. Recently, Liu et al.[13] presented the results of CL and Transmission Electron Microscopy (TEM) performed on the exactly same regions of thinned GaN/sapphire foils and gave clear evidence that basal plane stacking faults (BPSF) are responsible for an emission line at 3.41 eV, which is quite close to the peak observed at 3.39 eV for our 8° off-axis sample. *ii)* A prominent peak at 3.21 eV and its LO phonon replica at 3.12 eV and 3.03 eV. Very recently, Bai et al.[14] combined PL and TEM measurements and suggested that prismatic stacking faults and/or the associated stair-rod dislocations may be responsible for the PL features in the ~3.21 eV range. *iii)* A narrow feature at 3.28 eV. The best candidate for this peak is the exciton transition of the c-GaN inclusions observed by micro-Raman and TEM.

Figure 5 shows the CL spectra in the energy range of the donor bound exciton. As expected, this transition line decreases linearly with the biaxial stress. Using the coefficient $dE_A/d\sigma^B$ of the A exciton of Ref.[11], the DBE line shift permits to extrapolate the donor bound exciton transition energy for unstrained GaN at 3.47 eV, in good agreement with experimental data for free standing material. This confirms the tensile strain measured by X-Ray and Raman (away from the cracks).

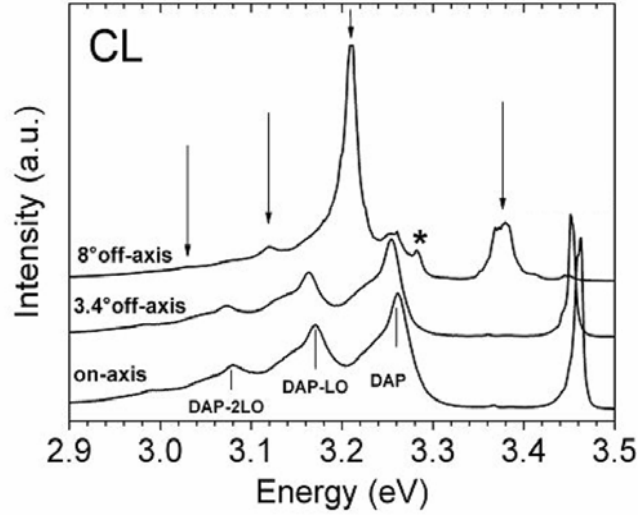

Fig.4

FIG. 4. Cathodoluminescence spectra measured at 12-20 K with $E_{EXC}$=20 keV of GaN epilayers deposited on misoriented 4H-SiC substrates (0°, 3.4° and 8° off-axis).

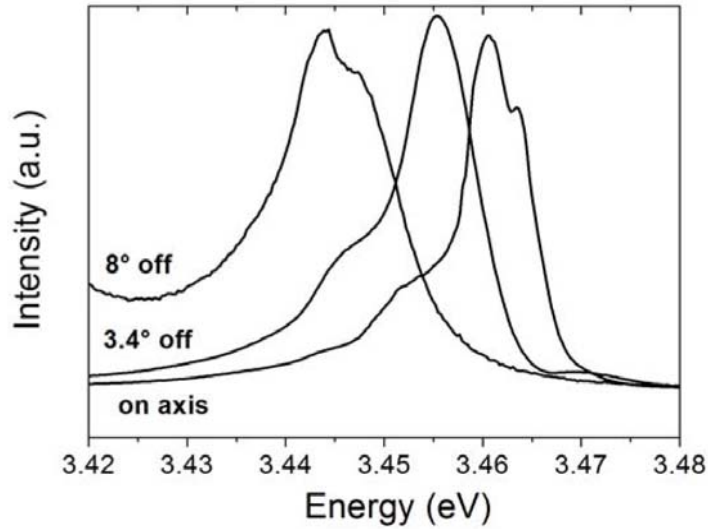

Fig.5

FIG. 5. Excitonic region of cathodoluminescence spectra measured at 12-20 K with $E_{EXC}$=20 keV of GaN epilayers deposited on misoriented 4H-SiC substrates (0°, 3.4° and 8° off-axis).

## IV. Discussion

The strain values measured in the three GaN layers grown during the same run on (0001) 4H-SiC substrates with different miscut angles (0°, 3.4° and 8° off [0001] towards [11-20] axis) show that the strain relaxation mechanisms involved are dependent of the substrate miscut angle. At a microscopic scale, the strain distribution of the on axis sample is homogeneous over the whole surface of the sample. For this sample, threading dislocations are formed upon coalescence of misoriented islands during the first step of the growth[15]. These isolated threading dislocations relaxes the strain through a plastic deformation yielding an uniformly and slightly stressed layer over the whole surface with a $R^B$ value far from the ideal elastic deformation case. The samples grown off axis are more favorable to a step flow growth, limiting the island formation and so decreasing the isolated threading dislocation density[5]. On the other hand, Rudziński et al. observed the formation of walls of dislocation along the [11-20] direction[10] for the 8° miscut angle. The observation of cubic inclusions[10] in sample #3 is certainly due to the anisotropic growth with triangularly shaped islands (2D nucleation growth mode)[5]. For the miscut samples, others relaxation mechanisms occur. Recently, Huang et al.[6] found that unpaired geometrical partial misfit dislocations (GPMDs) are formed along the step of the AlN/6H-SiC interface to correct the stacking sequence between 2H and 6H structures for misoriented substrates. This strain relaxation mechanism must be present at the AlGaN/SiC interface with a 2H/4H stacking structure in this case. For this system, the proportion of steps corresponding to GPMDs is 1/2 (1/3 for 2H with 6H [HuangPRL05]) since the number of cubic sites is 1/2 in the ABAC stacking sequence (one cubic site is defined as the B bilayer in the ABC sequence and one hexagonal site is defined as the B layer in the ABA sequence). This mechanism takes place during the step-flow growth, giving GPMDs lines along [1-100] axis and so, relaxing the [11-20] strains. Huang et al.[6], on the basis of geometrical considerations (miscut angle and stacking sequence) argue that the number of GPMDs can be controlled, relaxing more or less the [11-20] strains. This explains the difference between the samples #2 and #3. Indeed, the critical miscut angle for 2H-GaN/4H-SiC[16] is 6.5° which is closer for sample #3 than #2. During the growth and prior to cooling, more GPMDs are formed along the [1-100] axis for sample #3 than sample #2, limiting more efficiently the [11-20] strain. Then, during the cooling step of the growth, the thermal expansion coefficient difference between GaN and SiC puts the GaN epilayer under tensile in-plane strain, and then creates some cracks[17]. So, during the cooling step, the thermal expansion coefficient difference induces a high tensile strain along the [1-100] direction for the two miscut angles. This explains the formation of cracks along the [11-20] direction perpendicular to the [1-100] direction. The [11-20] strain is highly relaxed by the GPMDs for the sample #3 and not so much for sample #2, explaining the formation of additional cracks in the [2-1-10] and [-12-10] directions (not perpendicular to the [1-100] axis) for this sample.

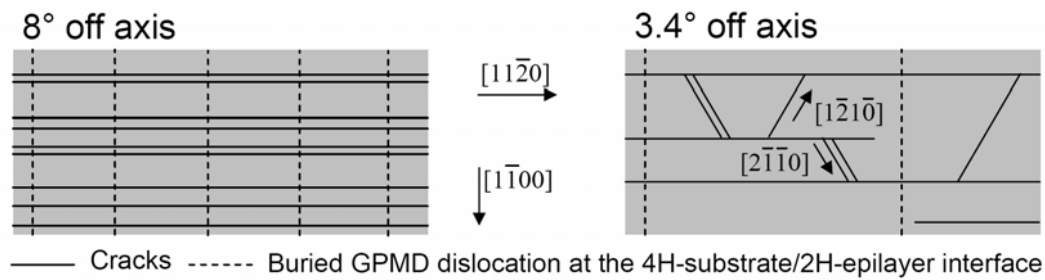

Fig.6

FIG. 6. Schematic draws of the crack and geometrical partial misfit dislocations (at the substrate epilayer interface) distribution in GaN grown on 4H-SiC substrates (3.4° and 8° off-axis).

This is schematically represented in Fig. 6 for samples #2 and #3. For the off cut samples (#2 and #3), the strain relaxation seems to be mainly governed by two successive steps:

i) Creation of GPMDs along the [1-100] axis[6], misfit dislocation walls along the [11-20] axis[10] and basal plane stacking faults, prismatic stacking faults and stair rod dislocation[18] during the step flow growth due to the misorientation of the substrate.

ii) Creation of cracks along the [11-20] direction (and also the [2-1-10] and [-12-10] for the sample #2) during the cooling process due to the difference of the thermal expansion coefficients and in particular directions to minimize the residual stress.

The resulting layers are heterogeneously in-plane strained with a strong plastic relaxation near the cracks and in between, an elastic deformation close to the ideal case with a lower threading isolated dislocation density than in the on axis sample.

## V. Conclusion

The strain of GaN layers grown by MOCVD on three vicinal 4H-SiC substrates (0, 3.4 and 8° offcut towards the [11-20] axis) have been measured and analyzed. For on axis samples, the threading dislocations due to the coalescence of misoriented islands during the first step of the growth relaxed the strain, resulting in a slightly but homogeneously tensile in-plane strained layer. For the miscut substrate, step flow growth occurs and other relaxation mechanisms appear during growth, such as the creation of various types of dislocations and stacking faults. During the cooling down step, cracks are formed in particular directions in order to limit the residual stress due to the difference of thermal expansion coefficients and to defect formation. Away from the cracks, the GaN areas are under a tensile in-plane stress with an almost pure elastic biaxial deformation.

## Acknowledgements

The financial support of M. Rudziński by Philips Semiconductors B.V. Nijmegen is gratefully acknowledged. J.P. and E.B. wish to thank F. Omn\`{e}s (Institut NEEL, CNRS, Grenoble, France) for fruitful discussions